\documentclass[acmsmall,screen]{acmart}
\usepackage{xcolor}
\usepackage{pifont}
\usepackage{booktabs}
\usepackage{graphicx}
\usepackage{tcolorbox}
\usepackage{multirow}
\usepackage{enumitem}
\usepackage{colortbl}
\usepackage{subcaption} 
\usepackage{xspace}
\usepackage{listings}
\usepackage{adjustbox}
\usepackage{xcolor}

\definecolor{mGreen}{rgb}{0,0.6,0}
\definecolor{mGray}{rgb}{0.5,0.5,0.5}
\definecolor{mPurple}{rgb}{0.58,0,0.82}
\definecolor{backgroundColour}{rgb}{0.95,0.95,0.92}
\newcommand{\lstbg}[3][0pt]{{\fboxsep#1\colorbox{#2}{\strut #3}}}
\definecolor{lightred}{RGB}{255,230,230}
\definecolor{lightgreen}{RGB}{230,255,230}

\newcommand{\find}[1]{
\begin{tcolorbox}[leftrule=1mm,toprule=0mm,bottomrule=0mm,left=1pt,right=2pt,top=2pt,bottom=2pt]
\em #1
\end{tcolorbox}
}

\lstdefinestyle{CStyle}{
    backgroundcolor=\color{backgroundColour},   
    commentstyle=\color{mGreen},
    keywordstyle=\color{magenta},
    numberstyle=\tiny\color{mGray},
    stringstyle=\color{mPurple},
    basicstyle=\footnotesize\ttfamily,
    breakatwhitespace=false,         
    breaklines=true,                 
    captionpos=b,                    
    keepspaces=true,                 
    numbers=left,                    
    numbersep=5pt,
    showspaces=false,                
    showstringspaces=false,
    showtabs=false,                  
    tabsize=2,
    language=C
}

\usepackage{cleveref}
\usepackage{listings}
\crefname{lstlisting}{listing}{listings}
\Crefname{lstlisting}{Listing}{Listings}
\usepackage{xcolor}

\definecolor{codegreen}{rgb}{0,0.6,0}
\definecolor{codegray}{rgb}{0.5,0.5,0.5}
\definecolor{codepurple}{rgb}{0.58,0,0.82}
\definecolor{shallowred}{rgb}{1,0.8,0.8}
\definecolor{verylightgray}{rgb}{0.97, 0.97,0.97}

\lstdefinestyle{mystyle}{
    language=C++, 
    commentstyle=\color{codegreen},
    keywordstyle=\color{blue},
    stringstyle=\color{codepurple},
    basicstyle=\ttfamily\scriptsize,
    breakatwhitespace=false,
    breaklines=true,              
    captionpos=b,                    
    keepspaces=true,                 
    numbers=left,                    
    numbersep=5pt,                  
    showspaces=false,                
    showstringspaces=false,
    frame=single,                    
    xleftmargin=0.05\linewidth,
    xrightmargin=0.05\linewidth,
    framexleftmargin=0.05mm,            
    framexrightmargin=0.05mm,           
    framextopmargin=0mm,             
    framexbottommargin=0mm,    
    showtabs=false,                  
    tabsize=2,
    escapeinside={(*@}{@*)},
}
\lstdefinelanguage{diff}{
	frame=single,
    breakatwhitespace=false,
    breaklines=true,
	basicstyle=\ttfamily\scriptsize, 
	morecomment=[f][\color{red}]{---}, 
	morecomment=[f][\color{codegreen}]{+++},
	morecomment=[f][\lstbg{red!20}]{-},
	morecomment=[f][\lstbg{green!20}]{+},
	morecomment=[f][\color{blue}]{@@},
    xleftmargin=0.03\linewidth,
    xrightmargin=0.005\linewidth,
    framexleftmargin=0.05mm,            
    framexrightmargin=0.0mm,           
    framextopmargin=0mm,             
    framexbottommargin=0mm,
    numbers=left,                    
}

\usepackage{threeparttable}
\AtBeginDocument{%
  }

\setcopyright{acmlicensed}
\copyrightyear{2025}
\acmYear{2025}
\acmDOI{XXXXXXX.XXXXXXX}

\acmConference[Conference acronym 'XX]{Make sure to enter the correct
  conference title from your rights confirmation emai}{Feb 24,
  2025}{Woodstock, NY}
\acmISBN{978-1-4503-XXXX-X/18/06}

\newcommand{\method}{{\textsc{VulCoCo}}\xspace}
\newcommand{\benchmark}{{\textsc{SyVC}}\xspace}

\definecolor{removedline}{RGB}{255,230,230}
\definecolor{addedline}{RGB}{230,255,230}

\begin{document}

\title{\method: A Simple Yet Effective Method for Detecting Vulnerable Code Clones}

\setcounter{page}{1}

\author{Tan Bui}
\affiliation{%
  \institution{Singapore Management University}
  \country{Singapore}
}
\email{ngoctanbui@smu.edu.sg}

\author{Yan Naing Tun}
\affiliation{%
  \institution{Singapore Management University}
  \country{Singapore}
}
\email{yannaingtun@smu.edu.sg}

\author{NGUYEN Phuc Thanh}
\affiliation{%
  \institution{Singapore Management University}
  \country{Singapore}
}
\email{ptnguyen@smu.edu.sg}

\author{Yindu Su}
\affiliation{%
  \institution{Singapore Management University}
  \country{Singapore}
}
\email{yindusu@smu.edu.sg}

\author{Ferdian THUNG}
\affiliation{%
  \institution{Singapore Management University}
  \country{Singapore}
}
\email{ferdianthung@smu.edu.sg}

\author{Yikun Li}
\affiliation{%
  \institution{Singapore Management University}
  \country{Singapore}
}
\email{yikunli@smu.edu.sg}

\author{Han Wei ANG}
\affiliation{%
  \institution{GovTech}
  \country{Singapore}
}
\email{ANG_Han_Wei@tech.gov.sg}

\author{Yide Yin}
\affiliation{%
  \institution{GovTech}
  \country{Singapore}
}
\email{YIN_Yide@tech.gov.sg}

\author{Frank Liauw}
\affiliation{%
  \institution{GovTech}
  \country{Singapore}
}
\email{Frank_LIAUW@tech.gov.sg}

\author{Shar Lwin Khin}
\affiliation{
\institution{Singapore Management University}
\country{Singapore}
}
\email{lkshar@smu.edu.sg}

\author{Ouh Eng Lieh}
\affiliation{
\institution{Singapore Management University}
\country{Singapore}
}
\email{elouh@smu.edu.sg}

\author{Ting Zhang}
\affiliation{%
 \institution{Singapore Management University}
 \country{Singapore}}
\email{tingzhang.2019@phdcs.smu.edu.sg}
\authornote{Ting Zhang is the corresponding author.}

\author{David Lo}
\affiliation{%
  \institution{Singapore Management University}
  \country{Singapore}
}
\email{davidlo@smu.edu.sg}

        
    


\renewcommand{\shortauthors}{Bui et al.}

\begin{abstract}
Code reuse is common in modern software development, but it can also spread vulnerabilities when developers unknowingly copy risky code.
The code fragments that preserve the logic of known vulnerabilities are known as vulnerable code clones (VCCs). 
Detecting those VCCs is a critical but challenging task. 
Existing VCC detection tools often rely on syntactic similarity or produce coarse vulnerability predictions without clear explanations, limiting their practical utility. 

In this paper, we propose \method, a lightweight and scalable approach that combines embedding-based retrieval with large language model (LLM) validation. Starting from a set of known vulnerable functions, we retrieve syntactically or semantically similar candidate functions from a large corpus and use an LLM to assess whether the candidates retain the vulnerability. 
Given that there is a lack of reproducible vulnerable code clone benchmarks, we first construct a synthetic benchmark that spans various clone types. 

Our experiments on the benchmark show that \method outperforms prior state-of-the-art methods in terms of precision@k and mean average precision (MAP). 
In addition, we also demonstrate \method's effectiveness in real-world projects by submitting 400 pull requests (PRs) to 284 open-source projects.
Among them, 75 PRs were merged, and 15 resulted in newly published CVEs. 
We also provide insights to inspire future work to further improve the precision of vulnerable code clone detection.

\end{abstract}

\maketitle

\section{Introduction}

Modern software development is increasingly reliant on open-source components and code reuse~\cite{kula2018developers, decan2018evolution}. While this accelerates development, it also facilitates the unintentional propagation of security vulnerabilities~\cite{li2016vulpecker, pashchenko2020vuln4real}. A particularly concerning problem
is code cloning, which is when a code fragment containing a vulnerability is copied across multiple codebases, potentially with superficial edits that make its origin harder to recognize~\cite{sajnani2016sourcerercc, nguyen2009clone}. These vulnerable clones can remain unnoticed for years, silently exposing downstream projects to security risks~\cite{kim2017vuddy, li2006cp}.

Detecting such vulnerable clones is challenging. Unlike generic code clone detection, which focuses on syntactic or semantic similarity~\cite{roy2009comparison, sajnani2016sourcerercc}, vulnerable code clone (VCC) detection must also consider the security status of the cloned code. A clone may look similar but be partially patched, altered in non-functional ways, or even fixed in subtle semantic forms that pass through surface-level similarity checks~\cite{bowman2020vgraph}. This means detecting similarity alone is not enough, we must also determine whether the clone still retains the vulnerable logic~\cite{xiao2020mvp,kim2017vuddy}.

Several prior studies have proposed new approaches to detect VCCs, including syntax-based~\cite{jang2012redebug, kim2017vuddy}, slice-based~\cite{alomari2025slicing}, and graph-based approaches~\cite{bowman2020vgraph}. 
However, these studies have two major limitations. 
First, some methods~\cite{kim2017vuddy, xiao2020mvp, woo2022movery} rely on strict similarity heuristics to generate a signature for each known vulnerability, which is then used to match against functions or code fragments in the target repositories.
Although these tools also adopt abstraction strategies to enhance generalizability, they still struggle to identify non-trivial cases such as Type-3 and Type-4 clones. 
They often fail when the clone has small edits or follows a different coding style or structure, causing them to miss clones that retain the vulnerable behavior but differ in syntax or control flow.
Our investigation shows that while some existing tools achieve high precision, they frequently miss clones that are semantically similar but structurally different from known vulnerable code.
Second, many of these studies rely on manual, ad-hoc, and small benchmarks, limiting their reproducibility and scalability~\cite{alomari2025slicing, xiao2020mvp, feng2024fire}. 
Such practice suggests that each time a new method is proposed, there is a constant need for another round of manual evaluation.

To tackle the first limitation, we propose \method, a simple yet effective method for scalable VCC detection. 
Our approach combines semantic embedding-based retrieval to identify candidate clones and LLM validation to assess whether retrieved functions still contain the vulnerable behavior. 
To address the second limitation, we introduce \benchmark (\underline{Sy}nthetic \underline{V}ulnerable \underline{C}lones), a retrieval-based synthetic benchmark that contains a wide variety of realistic clones. 
The benchmark is built from 100 real-world vulnerable-fixed function pairs, selected from CVEs spanning the CWEs in the 2024 CWE Top 25 Most Dangerous Software
Weaknesses.\footnote{\url{https://cwe.mitre.org/top25/archive/2024/2024\_cwe\_top25.html}} 
For each function, we generate 5
clones reflecting the four standard types of code similarity~\cite{roy2007survey},
including exact copies, identifier substitutions, structural edits, and semantically equivalent rewrites. These perturbations are crafted using a combination of manual heuristics and LLM generation to ensure realism and diversity. Each clone is labeled according to whether it originates from a vulnerable or fixed function, resulting in a balanced set of 1,000 labeled clone candidates. To ensure the quality of \benchmark, we first randomly sampled 100 synthesized functions, then two authors independently manually checked if they were actually cloned from the vulnerable or the fixed function, and whether the security status (vulnerable or fixed) still remains. Our manual verification confirms that all LLM-generated clones are indeed clones of the vulnerable/fixed source function and retain the same vulnerability status, although Type-4 clones remain less straightforward.
\benchmark provides a standardized benchmark for evaluating the precision and robustness of vulnerable clone detection techniques under realistic and diverse code transformations. 

Beyond the benchmark, we also apply our method to real-world open-source repositories. More specifically, we ran \method on the top 5,000 most-starred
repositories in C, C++, or Java on GitHub while discarding inactive ones. 
We have submitted 400 
PRs to patch detected vulnerable clones in 284 projects, of which 75 have been merged into the upstream main branches. 
In addition, 15 CVEs have been published as a direct result of these discoveries. These results confirm the practical utility of our approach and demonstrate its potential to enhance vulnerability detection and mitigation at scale.

Our contributions can be summarized as follows:
\begin{itemize}[leftmargin=*]
    \item We construct \benchmark, a synthetic, retrieval-based benchmark for VCC detection, which aims to enable reproducible evaluation of VCC detection methods.
    \item We propose a lightweight technique named \method, which combines embedding-based retrieval with LLM-based vulnerability reasoning. \method outperforms prior state-of-the-art methods, achieving up to 119\% higher MAP on the \benchmark benchmark. Our method can also find more than 7 times as many VCCs compared to the prior state-of-the-art method FIRE~\cite{feng2024fire}, while keeping good precision when tested on real software projects.
    \item We demonstrate the real-world effectiveness of \method by submitting 400 PRs, with 75 merged and 15 new CVEs published to date.
\end{itemize}

The remaining part of this paper is organized as follows: Section~\ref{sec:background} discusses the background and motivation of this work.
Section~\ref{sec:methodology} elaborates on our proposed framework \method.
In Section~\ref{sec:setup}, we provide details about the experimental setup.
Section~\ref{sec:results} presents experimental results.
We also discuss threats to validity in Section~\ref{sec:discussion}.
Section~\ref{sec:related} talks about related work.
We conclude our work and discuss the future work in Section~\ref{sec:conclusion}.

\section{Background and Motivation}
\label{sec:background}

In this section, we first review the concepts most relevant to VCC, and then discuss the motivation of our study.

\subsection{Definition of Vulnerable Code Clone (VCC)}


Code clones have been a well-established research topic for more than two decades~\cite{kamiya2002ccfinder,kim2005empirical}, and they are commonly classified into four types based on their level of similarity~\cite{roy2007survey}.

\begin{itemize}[leftmargin=*]
    \item Type-1: The two code snippets are exact copies of each other except for whitespace and comments.
    \item Type-2: The two code snippets are syntactically similar except for changes in identifiers, literals, types, and superficial changes as in Type-1.
    \item Type-3: The two code snippets belong to Type-2 clones and they have some further modifications. Some statements can be modified, added to, or removed from one code snippet.
    \item Type-4: The two code snippets are semantically similar; they perform the same functionality but are syntactically different.
\end{itemize}

As a specific variant of code clone, VCC also inherits these types.
The categorization of clone types helps measure the difficulty of detecting VCCs. 
As we move from Type-1 to Type-4, the syntactic resemblance between clones decreases, making automated detection increasingly challenging. 
However, vulnerable logic can still exist across all clone types.

Unlike general-purpose clone detection, VCC detection requires more than identifying similarity. 
For two functions to constitute a VCC pair, both must still be vulnerable and share the \emph{same} vulnerability.
By contrast, a vulnerable function and its patched version can still be general code clones, yet only the original is security‑relevant, and thus the pair is not a VCC.
Formally, functions $f_1$ and $f_2$ are a VCC pair if (1) each remains vulnerable, (2) both expose the same vulnerability, and (3) they fall into one of the four clone categories defined above.

We demonstrate this key distinction using a vulnerable function and its fixed version, showing how they form a general code clone but not a VCC since only one remains vulnerable.
A common mistake is incorrectly computing the midpoint of an array range, leading to an integer overflow vulnerability.
Although two versions of the code may be syntactically or semantically similar, only one is unsafe. 
In CVE-2014-9666\footnote{\url{https://nvd.nist.gov/vuln/detail/CVE-2014-9666}}, as shown in Listing~\ref{lst:example}, the vulnerable version checks \texttt{decoder->strike\_index\_array + 8 * decoder -> strike\_index\_count > face->sbit\_table\_size}, which may overflow when \texttt{decoder->strike\_index\_array + 8 * decoder->strike\_index\_count} exceeds the maximum integer limit. The fixed version avoids this by changing the checks to \texttt{decoder->strike\_index\_count > ( face->sbit\_table\_size - decoder-> strike\_index\_array ) / 8}.

\begin{lstlisting}[language=diff,caption={Patch in \texttt{mem\_check\_range}}, label={lst:example}, float=t]
static FT_Error
  tt_sbit_decoder_init( TT_SBitDecoder       decoder, ... )
  {
diff --git a/src/sfnt/ttsbit.c b/src/sfnt/ttsbit.c
index da6b01ba4..b37bd7dbb 100644
--- a/src/sfnt/ttsbit.c
+++ b/src/sfnt/ttsbit.c
@@ -394,9 +394,11 @@
      p                          += 34;
      decoder->bit_depth          = *p;
-     if ( decoder->strike_index_array > face->sbit_table_size             ||
-          decoder->strike_index_array + 8 * decoder->strike_index_count >
-            face->sbit_table_size                                         )
+     /* decoder->strike_index_array +                               */
+     /*   8 * decoder->strike_index_count > face->sbit_table_size ? */
+     if ( decoder->strike_index_array > face->sbit_table_size           ||
+          decoder->strike_index_count >
+            ( face->sbit_table_size - decoder->strike_index_array ) / 8 )
        error = FT_THROW( Invalid_File_Format );
    }
    ...
}
\end{lstlisting}

\subsection{Problem Formulation}
Most existing studies evaluate VCC detection tools using classification metrics~\cite{alomari2025slicing, bowman2020vgraph, kim2017vuddy, jang2012redebug, xiao2020mvp}. Typically, they run the tools on open-source repositories and manually validate the results. The ground-truth set is formed by collecting all clones identified by the tools considered in each study and label them as true/false clones. This set is then used to compute Precision, Recall, and F1 Score.
However, we argue that the nature of the task is better formulated as a code retrieval problem, where the goal is to find all the code snippets that match a known vulnerable example, rather than only making binary decisions about pairs of code snippets.

Given a known vulnerable function (i.e., the query), the objective is to retrieve code snippets from a large corpus that are likely to contain semantically similar and potentially vulnerable clones. 
This framework aligns with the search strategy adopted by many studies, such as VulPecker~\cite{li2016vulpecker} or ReDeBug~\cite{jang2012redebug}.

Formally, let $v$ be a known vulnerable function, and let $T = \{t_1, t_2, \ldots, t_n\}$ be a collection of candidate target functions. 
The task is to return a ranked list $R(v) \subseteq T$, where higher-ranked functions are more likely to \textit{suffer from the same vulnerability} as $v$. 


\subsection{Motivation}
\subsubsection{Motivating Example: Missed Clones in Practice}

To demonstrate the limitations of existing tools and the advantages of our approach, consider the following example. 
Listing~\ref{lst:org} shows a function containing a known vulnerability from \texttt{file/file}\footnote{\url{https://github.com/file/file}} project, and was reported as CVE-2015-8865.~\footnote{\url{https://nvd.nist.gov/vuln/detail/cve-2015-8865}} It reallocates memory for an internal structure without checking whether the allocation size may overflow, which can lead to denial of service or execution of arbitrary code via a crafted input.

\begin{lstlisting}[caption={Original vulnerable file\_check\_mem function}, label=lst:org,float=t]
protected int
file_check_mem(struct magic_set *ms, unsigned int level) {
    size_t len;
    if (level >= ms->c.len) {
        len = (ms->c.len += 20) * sizeof(*ms->c.li);
        ms->c.li = CAST(struct level_info *, (ms->c.li == NULL) ?
            malloc(len) : realloc(ms->c.li, len));
        if (ms->c.li == NULL) {
            file_oomem(ms, len);
            return -1;
        }
    }
    ...
    return 0;
}
\end{lstlisting}

A vulnerable clone of this function was found by our technique in \texttt{radareorg/radare2}~\footnote{\url{https://github.com/radareorg/radare2}}, with changes to formatting, function naming, and style, but it still retains the core vulnerability. 
The clone function is shown in Listing~\ref{lst:clone}.

\begin{lstlisting}[caption={Clone vulnerable \_\_magic\_file\_check\_mem function}, label=lst:clone,float=t]
int __magic_file_check_mem(RMagic *ms, unsigned int level) {
	if (level >= ms->c.len) {
		size_t len = (ms->c.len += 20) * sizeof (*ms->c.li);
		ms->c.li = (!ms->c.li) ? malloc (len) :
		    realloc (ms->c.li, len);
		if (!ms->c.li) {
			__magic_file_oomem (ms, len);
			return -1;
		}
	}
	ms->c.li[level].got_match = 0;
	ms->c.li[level].last_match = 0;
	ms->c.li[level].last_cond = COND_NONE;
	return 0;
}

\end{lstlisting}

Although the vulnerable logic is clearly preserved, especially the unchecked increment and reallocation, existing tools failed to detect this clone due to subtle edits like variable renaming and formatting changes.
These changes placed the function just beyond the reach of exact-match or syntax-heavy tools like ReDeBug or MVP~\cite{jang2012redebug,xiao2020mvp}.

In contrast, our approach detects this clone during the embedding-based retrieval phase, recognizing its syntactic resemblance. Rather than discarding it, we forward it to an LLM for further analysis. The LLM then validates that the vulnerable logic remains, allowing us to correctly identify the VCC.

This example highlights a key strength of our method: rather than relying solely on strict matching rules, we adopt a more flexible retrieval approach followed by an LLM validation step. 
This enables us to retain potentially vulnerable candidates that other tools overlook, thereby improving both recall and real-world utility.

\subsubsection{Reproducibility Challenges}

Another limitation of existing VCC detection tools is that they typically rely on manual result inspection~\cite{alomari2025slicing, jang2012redebug, kim2017vuddy} or ad hoc datasets~\cite{bowman2020vgraph}, making it difficult to scale or reproduce. 
To address this, we construct a synthetic benchmark of 100 pairs of vulnerable-fixed function, each seeded with a known vulnerability. 
We generate Type-1 clones using common superficial edits (e.g., whitespace or comment changes), and leverage Claude Sonnet 4~\cite{claude}
to produce more challenging Type-2, Type-3, and Type-4 clones. These include identifier renaming, statement reordering, and semantically equivalent but syntactically different implementations. This dataset enables reproducible evaluation of vulnerable clone detectors across a wide range of realistic clone types.


Beyond benchmarks, we validate our system on real-world open-source projects by submitting PRs with automatically generated fixes for detected clones. 
Until June 7th, 2025, 75 of these have been accepted, demonstrating the practical value of our method and its potential to enhance vulnerability detection and mitigation in the open-source ecosystem.

\section{Methodology}
\label{sec:methodology}

\begin{figure}[t]
    \centering
    \includegraphics[width=\linewidth]{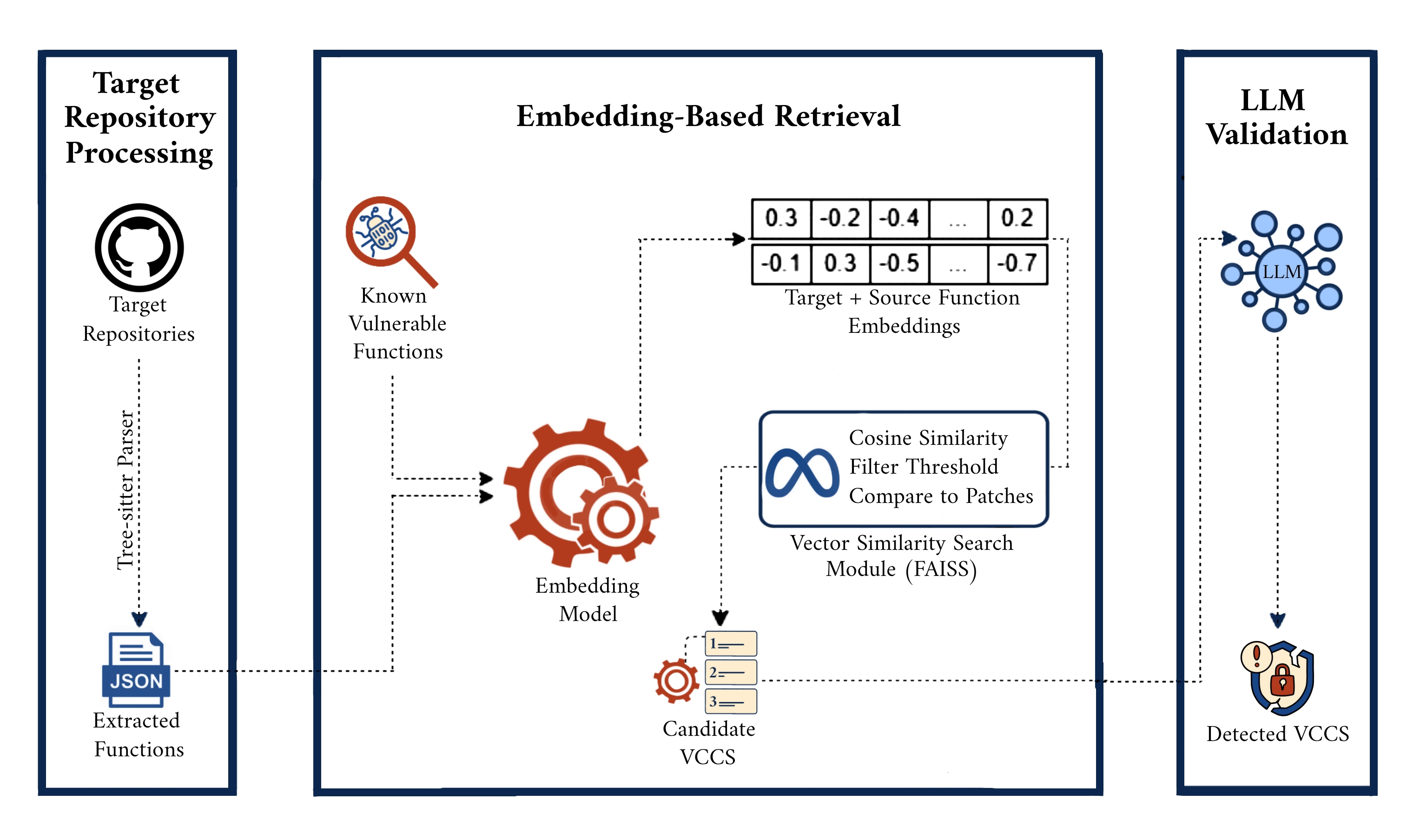}
    \caption{Overview of the system pipeline.}
    \label{fig:system_pipeline}
\end{figure}

We propose a three-phase pipeline for detecting VCCs at the function level. The overview of \method is shown in Figure~\ref{fig:system_pipeline}. 
The pipeline begins with target repository processing to extract functions, followed by embedding-based retrieval that uses cosine similarity to identify candidate VCCs, and concludes with LLM validation to confirm true vulnerable code clones.

\subsection{Target Repository Processing}

To identify potential VCCs, we collect and process target repositories through two main steps:

\begin{enumerate}[leftmargin=*]
    \item \textbf{Repository Collection}: We systematically collect a curated set of source code repositories from GitHub across multiple programming languages, prioritizing popular and actively maintained projects. 
    We select open-source repositories in C, C++, or Java from the top 5000 most-starred projects in each language on GitHub. We then apply two filtering criteria: (1) repositories must have commits within the last 10 months to ensure they remain active, and (2) repositories must have a pull request merge rate above 10\% to indicate openness to external contributions.
    \item \textbf{Function-Level Parsing}: We extract function-level units from the collected repositories using \texttt{tree-sitter}.\footnote{\url{https://tree-sitter.github.io/tree-sitter/}} This parsing process supports multiple languages (e.g. Java, C/C++) via language-specific grammar bindings. Functions are extracted based on node types (e.g., \texttt{method\_declaration}, \texttt{function\_declaration}), and are stored in JSON format for downstream processing.
    
\end{enumerate}

\subsection{Embedding-Based Retrieval}

To find semantic code clones, we use an embedding-based approach that compares target functions to a set of known vulnerable source functions. We obtain the vulnerable source functions from the standardized vulnerable-fixed function pairs from existing datasets such as CleanVul~\cite{li2024cleanvul} and PrimeVul~\cite{ding2024vulnerability}. They serve as reference queries in the retrieval phase.

\vspace{4px}
    \noindent\textbf{Code Embedding}: Prior to embedding, we apply a series of preprocessing steps to source and target functions: comments are removed, whitespace is standardized (e.g., consistent indentation and spacing), and keyword casing is normalized. These steps help reduce superficial differences and improve the quality of semantic matching. We then encode the preprocessed functions using the \texttt{jina-embeddings-v4}~\cite{günther2025jinaembeddingsv4universalembeddingsmultimodal} model, which captures the code semantic. We select \texttt{jina-embeddings-v4} for its task-specific adapters optimized for code-related tasks and its multilingual support, which is essential for handling C, C++, and Java code.
    
    \vspace{4px}
    \noindent\textbf{Similarity Search}: For each target function, we identify its most similar vulnerable source function
    using inner product-based search on L2-normalized embeddings (i.e., equivalent to cosine similarity). We retrieve all candidate functions from the FAISS~\cite{johnson2019billion, douze2024faiss} index whose similarity exceeds a threshold $t$, thereby selecting high-confidence matches. We employ FAISS for its optimized performance on high-dimensional vector similarity search, allowing efficient retrieval across our large-scale repository collection.
    
    \vspace{4px}
    \noindent\textbf{Clone Filtering}: For each candidate, we calculate its cosine similarity to the fixed source function.
    If the candidate appears closer to the fixed version than to the vulnerable one, it is discarded under the assumption that it likely reflects the patched logic. 
    This comparative filtering addresses a key challenge in VCC detection: distinguishing between truly vulnerable clones and code that implements similar functionality but includes security fixes, as both may have high similarity to the original vulnerable function.
    

This combination of embedding and similarity-based filtering yields a set of candidate vulnerable clones for further inspection.

\subsection{LLM-Based Validation}

To increase precision, we apply an LLM-based validation step to further validate if the candidates are actually VCCs. 
Each candidate clone is sent to Claude Sonnet 4~\cite{claude} via API with a structured prompt that includes the original vulnerable function and its fixed version. 
We select Claude Sonnet 4 for its strong performance in structured reasoning and code understanding, as demonstrated in Anthropic’s system evaluations across software development and security-oriented tasks~\cite{claudesystemcard}.
Its ability to follow complex instructions, analyze code logic, and reason across extended contexts makes it suitable for the task of VCC detection.
The model decides whether the candidate still has the vulnerability. The prompt template we use is shown in Listing~\ref{lst:prompt_template}.

\begin{lstlisting}[caption={Prompt template for LLM-based validation}, label={lst:prompt_template},float=t]
PROMPT_TEMPLATE = """
Analyze the following code functions to determine if the cloned function contains the same vulnerability as the original function:

ORIGINAL FUNCTION (Known vulnerable):
{original_function}

FIXED FUNCTION (Patched version):
{fixed_function}

CLONED FUNCTION (To be assessed):
{cloned_function}

Please assess whether the cloned function is vulnerable to the same issue that was fixed in the original function.
NOTE: A function that merely CALLS the original vulnerable function should NOT be considered a vulnerable clone unless it also IMPLEMENTS similar vulnerable logic itself. Focus on whether the cloned function contains similar vulnerable code patterns, not just whether it uses the vulnerable function.
IMPORTANT: Respond ONLY with valid JSON in the exact format below. Do not include any explanatory text before or after the JSON.

=== JSON RESPONSE ===
{{
  "is_vulnerable": true/false,
  "confidence_level": 1-5,
  "justification": "Detailed explanation of why the clone is or is not vulnerable. For vulnerable cases, explain what specific vulnerability pattern is present. For non-vulnerable cases, explain what protections/fixes are already in place that prevent the vulnerability."
}}
=== END JSON ===
"""
\end{lstlisting}

This semantic validation helps differentiate between true vulnerable clones and non-vulnerable or partially fixed variants that may still resemble the original code. In addition to binary labels (vulnerable or not), the LLM validation component also generates brief justifications for its decisions. These explanations offer insights into why a candidate was flagged as a potentially vulnerable, which can help developers and security analysts better assess the results.
The NOTE instruction reduces false positives by preventing the LLM from wrongly labeling functions that simply accidentally call the vulnerable function as vulnerable clones. The IMPORTANT instruction ensures JSON-only responses for easy automated post-processing.

\section{Experimental Setup}
\label{sec:setup}

In this section, we first introduce the research questions (RQs) we plan to answer.
Second, we outline the datasets that we use to evaluate our method \method.
Next, we describe the baselines, implementation details, and evaluation metrics.

\subsection{Research Questions}
\begin{itemize}[leftmargin=*]
    \item{\textbf{RQ1:} \textit{How effective and efficient is \method in detecting VCCs?}} \\
    In this RQ, we compare \method with prior state-of-the-art VCC detection methods.
    \item{\textbf{RQ2:} \textit{How do different components affect \method?}} \\
    In this RQ, we conduct an ablation study to understand the contribution of each component in our framework. In particular, we measure the impact of (1) LLM validation, (2) retrieval embedding model, and (3) similarity threshold on benchmark performance.
    \item{\textbf{RQ3:} \textit{How effective is \method in the real world scenario?}} \\
    In this RQ, we conduct two sets of evaluation: (1) We run \method and the baselines on real-world projects, and two of the authors manually verify the top 5 predictions by each method to get a ground-truth set. Later, with this set, we can compute the precision, recall, and F1 of each method. (2) We take a step further by submitting fix PRs for \method's detected clones.
\end{itemize}

\subsection{Datasets}
\subsubsection{Source Functions}
We took source functions from vulnerable/fixed function pairs collected from CleanVul~\cite{li2024cleanvul} and PrimeVul~\cite{ding2024vulnerability}, which provide high-quality examples of real-world vulnerabilities and their corresponding patches. 
To improve evaluation quality, we discard the top and bottom 5\% of functions based on the number of lines. 
Short functions are often ambiguous when cloned~\cite{kamiya2002ccfinder, roy2007survey, bazrafshan2013empirical}, and very long functions exceed the context limits of most LLMs. 
After filtering, we retain a total of 9,539 pairs of vulnerable and fixed functions to generate queries.

\subsubsection{Target Repositories}
We selected a total of 10 target repositories based on the following criteria:

\begin{itemize}[leftmargin=*]
    \item \textbf{Programming Language:} We focused on open source C/C++ and Java projects, as many security-related studies focus on these languages~\cite{woo2023v1scan, bui2024javavfc, bui2022vul4j}, and they are well supported by existing vulnerability datasets such as CleanVul~\cite{li2024cleanvul} and PrimeVul~\cite{ding2024vulnerability}, making them ideal for evaluating our method.
    \item \textbf{Repository Activity:} To ensure that we interact with active maintainers and have a reasonable chance of contributing fixes, we selected repositories with at least one commit in the last 10 months and a PR acceptance rate of at least 10\%.
    \item \textbf{Domain Diversity:} To assess the generalizability of our method, we selected repositories that span a range of application domains.
\end{itemize}

Table~\ref{tab:repo-overview} summarizes the selected repositories along with their programming languages, numbers of stars on GitHub, and domains.

\begin{table}[h]
\centering
\small
\caption{Overview of Selected Target Repositories}
\begin{tabular}{@{}lllll@{}}
\toprule
\multirow{2}{*}{\textbf{Repository}} & \multirow{2}{*}{\textbf{Language}} & \multirow{2}{*}{\textbf{\# Stars}} & \textbf{\# Parsed} & \multirow{2}{*}{\textbf{Domain}} \\
 &  &  & \textbf{Functions} & \\
\midrule
\rowcolor[HTML]{EFEFEF} 
aseprite/aseprite & C++ & 32,335 & 9,612 & Graphics/UI Tools \\
ClickHouse/ClickHouse & C++ & 40,958 & 45,392 & Database Management \\
\rowcolor[HTML]{EFEFEF} 
apache/flink & Java & 24,915 & 109,319 & Stream Processing Framework \\
jeecgboot/JeecgBoot & Java & 42,888 & 4,294 & Low-code AI Platforms \\
\rowcolor[HTML]{EFEFEF} 
coolsnowwolf/lede & C & 30,717 & 31,342 & Networking Software \\
micropython/micropython & C & 20,431 & 10,158 & Embedded System \\
\rowcolor[HTML]{EFEFEF} 
gentilkiwi/mimikatz & C & 20,240 & 5,590 & Security Framework \\
radareorg/radare2 & C & 21,748 & 20,190 & Reverse Engineering Framework \\
\rowcolor[HTML]{EFEFEF} 
valkey-io/valkey & C & 21,732 & 9,262 & Database Management \\
kingToolbox/WindTerm & C & 26,486 & 3,308 & Cross-Platform Terminal \\
\bottomrule
\end{tabular}
\label{tab:repo-overview}
\end{table}

\subsubsection{Synthetic Benchmark}

To support controlled and reproducible evaluation, we construct a synthetic benchmark from 100 
vulnerable-fixed function pairs sampled from the source functions described above. Each pair includes a vulnerable function and its corresponding fixed version.

For each of the 200 functions, which include 100 vulnerable functions and 100 fixed functions,
we generate five clones, consisting of:
\begin{itemize}[leftmargin=*]
    \item One Type-1 clone (whitespace and comment changes),
    \item One Type-2 clone (identifier and literal substitutions),
    \item Two Type-3 clones (statement reordering, insertions, or deletions), and
    \item One Type-4 clone (semantically equivalent rewrite).
\end{itemize}
This results in a total of 1,000 synthetic clones. We generate two Type-3 clones because Type-3 clones preserve enough structural similarity to be detectable by \method and the baselines, while still introducing meaningful variation that challenges detection methods. 
Type-1 clones are generated using simple automated transformations. Types 2–4 are synthesized using Claude Sonnet 4~\cite{claude} to ensure realistic and diverse perturbations.
Each clone inherits the vulnerability label of its origin (i.e., 500 clones are derived from vulnerable functions, and 500 from fixed functions). 
To ensure balanced representation, we sample at most 10 pairs of functions per CWE, focusing on high-impact vulnerability types. 
All selected CWEs belong to the 2024 CWE Top 25 Most Dangerous Software Weaknesses.\footnote{\url{https://cwe.mitre.org/top25/archive/2024/2024\_cwe\_top25.html}} 
To validate the quality of generated clones, we randomly sample 100 generated functions and manually verify that all are true clones of the correct clone type and the vulnerability status remains unchanged. Our manual checking process shows that all samples maintain their vulnerabilities and align with the assigned clone types, while Type-4 clones are less obviously related to the original function than other clone types.

Table~\ref{tab:clone-stats} summarizes the number and characteristics of each clone type, while Table~\ref{tab:cwe-distribution} shows the CWE distribution among source pairs. 

\begin{table}[h]
\centering
\caption{Statistics of Generated Synthetic Clones}
\small
\begin{tabular}{lrrrr}
\toprule
\textbf{Clone Type} & \textbf{\# Clones} & \textbf{\# Pos} & \textbf{\# Neg} & \textbf{Avg. |Token Diff|} \\
\midrule
\rowcolor[HTML]{EFEFEF} 
Type-1 (Whitespace/Comments)  & 200 & 100 & 100 & 15.79 \\
Type-2 (Identifiers/Literals) & 200 & 100 & 100 & 9.99 \\
\rowcolor[HTML]{EFEFEF} 
Type-3 (Statement Edits)      & 400 & 200 & 200 & 57.42 \\
Type-4 (Semantic Rewrite)     & 200 & 100 & 100 & 64.68 \\
\midrule
\textbf{Total}                & 1000 & 500 & 500 & -- \\
\bottomrule
\end{tabular}
\label{tab:clone-stats}
\end{table}

\begin{table}[h]
\centering
\caption{Distribution of CWE Types in Source Function Pairs}
\small
\begin{tabular}{lr}
\toprule
\textbf{CWE ID} & \textbf{\# Function Pairs} \\
\midrule
\rowcolor[HTML]{EFEFEF} 
CWE-79 (Cross-site Scripting)             & 10 \\
CWE-787 (Out-of-bounds Write)             & 10 \\
\rowcolor[HTML]{EFEFEF} 
CWE-22 (Path Traversal)                   & 10 \\
CWE-125 (Out-of-bounds Read)              & 10 \\
\rowcolor[HTML]{EFEFEF} 
CWE-78 (OS Command Injection)             & 10 \\
CWE-416 (Use After Free)                  & 10 \\
\rowcolor[HTML]{EFEFEF} 
CWE-94 (Code Injection)                   & 10 \\
CWE-20 (Improper Input Validation)        & 10 \\
\rowcolor[HTML]{EFEFEF} 
CWE-862 (Missing Authorization)           & 8  \\
CWE-77 (Command Injection)                & 6  \\
\rowcolor[HTML]{EFEFEF} 
CWE-89 (SQL Injection)                    & 3  \\
CWE-287 (Improper Authentication)         & 3  \\
\midrule
\textbf{Total}                            & 100 \\
\bottomrule
\end{tabular}
\label{tab:cwe-distribution}
\end{table}

Type-1 clones still have a fairly high average token difference of 15.79 with the original function, mostly because long comments are split into many tokens.
Type-2 clones change less and usually keep the same structure, so their average token difference is lower at 9.99.
Type-3 and Type-4 clones show much larger differences, with averages of 57.42 and 64.68 tokens.
This makes sense because they involve more changes like reordering, adding, or rewriting code.
These results show that our benchmark includes a wide range of clone types, from small edits to major rewrites that preserve the code semantics.


\subsection{Baselines}

To benchmark the effectiveness of our method, we compare it against a set of state-of-the-art VCC detection approaches:

\begin{itemize}[leftmargin=*]
    \item \textbf{Hash-Based Baseline}: This simple function-matching method abstracts names and parameters before computing hashes of the code. If a candidate has the same hash as a known vulnerable one, it is flagged. This gives us a basic point of comparison.
    \item \textbf{ReDeBug}~\cite{jang2012redebug}: 
    ReDeBug is a signature-based method that looks at lines removed during a vulnerability fix and the surrounding context, then uses them as a pattern to find similar unfixed code in other places.
    \item \textbf{MOVERY}~\cite{woo2022movery}: This method detects vulnerable code reuse by combining semantic patch analysis and dependency-aware filtering. It extracts essential and dependent lines from both the vulnerable and patched versions of a function, and identifies targets that include all vulnerable-related lines but exclude all patch-related ones. Syntax similarity is also used to confirm structural resemblance to known vulnerabilities.
    \item \textbf{FIRE}~\cite{feng2024fire}: FIRE is a multi-phase pipeline that first applies token and AST similarity filtering, then performs taint analysis to confirm whether the candidate function preserves the same data flow patterns as the original vulnerability.
    \item \textbf{SrcVul}~\cite{alomari2025slicing}: A slicing-based system that focuses on vulnerability-relevant variables. It computes semantic slice vectors and uses locality-sensitive hashing to identify structurally modified but semantically similar clones.
 
\end{itemize}

These baselines use different ways to find clones, from simple text matching to deeper analysis, allowing us to assess how our method performs relative to traditional tools in detecting transformed or subtle vulnerable code clones.

\subsection{Implementation Details}
Among the baselines, we re-implement MOVERY~\cite{woo2022movery} ourselves, as its original replication package lacks the vulnerability signature generation module, a critical component for its clone detection logic. Our implementation faithfully follows the paper's described workflow.
In addition, we implement a simple hash-based baseline for comparison. This technique uses normalized and abstracted code representations, where function parameters are replaced with FPARAM, local variables with LVAR, data types with DTYPE, and function calls with FUNCCALL.
All whitespaces and comments are then removed before applying the MD5 hash to identify potential clones. 
While not derived from prior publications, it serves as a simple and easy-to-understand baseline in our evaluation.
All other baselines, including ReDeBug~\cite{jang2012redebug}, FIRE~\cite{feng2024fire}, and SrcVul~\cite{alomari2025slicing}, are obtained from their official repositories.

To promote reproducibility and account for variation in LLM responses, we run the LLM-based validation stage three times on the benchmark using temperature 0 and consistent settings. 
We report the average results, with observed fluctuations in metrics staying below 1\%, indicating stable and consistent performance.

We also construct a validation set to guide parameter tuning and prompt design. 
This set consists of 100 synthetic clones generated from 10 vulnerable-fixed function pairs, covering all four clone types. 
It mirrors the structure of the main benchmark while being smaller in scale, ensuring minimal bias during evaluation.

\subsection{Evaluation Metrics}
\subsubsection{Synthetic Benchmark Evaluation}
On the synthetic benchmark, where ground-truth clone labels are available, we evaluate all baselines and our approach using ranking-based metrics: Precision@k (P@k) and MAP. These metrics measure the accuracy of the top-$k$ retrieved candidates for each vulnerable source function
. Since our benchmark includes exactly five ground-truth clones per query,
Recall@k is proportional to Precision@k, and F1@k adds no meaningful information. Therefore, we report only Precision@k and MAP.

Formally, for a given vulnerable source function:
\begin{align*}
\text{Precision@}k &= \frac{1}{|Q|} \sum_{q \in Q} \frac{|\text{top}_k(q) \cap G_q|}{k} \\
\text{AP}(q) &= \frac{1}{|G_q|} \sum_{i=1}^{N_q} \text{Precision@}i(q) \cdot \text{rel}_q(i) \\
\text{MAP} &= \frac{1}{|Q|} \sum_{q \in Q} \text{AP}(q)
\end{align*}

Where:
\begin{itemize}
    \item $|Q|$ is the number of queries (vulnerable source functions).
    \item $G_q$ is the set of ground-truth clones for query $q$ (with $|G_q| = 5$).
    \item $\text{top}_k(q)$ is the set of top-$k$ retrieved results for query $q$.
    \item $\text{rel}_q(i)$ is 1 if the $i$-th ranked result is in $G_q$, 0 otherwise.
\end{itemize}

Together, these metrics quantify the effectiveness of \method and allow us to compare with the baselines. Since Precision@k already reflects retrieval accuracy under our fixed ground-truth setup, we report it as the primary ranking metric. We additionally include MAP to capture overall ranking quality by rewarding systems that rank true positives higher in the candidate list.


\subsubsection{Real-World Evaluation}

In real-world scenarios, ground-truth clone labels are unavailable. 
We follow the evaluation protocol of prior work~\cite{xiao2020mvp,jang2012redebug} and assess LLM validation using manual inspection.
Specifically, we gather all predicted vulnerable clones from our method and baselines, and manually inspect each one to determine whether it truly retains the same vulnerability as the original source function. Two independent annotators performed this review, each holding at least a Bachelor's degree in Computer Science and 5+ years of programming experience. 
The final set, used as the ground truth, consists of manually labelled clones from all tools.
This set allows us to compute precision, recall, and F1-score for each method, ensuring fair and meaningful comparison.

\subsubsection{Real-World Case Studies}

Beyond validation accuracy, we conduct case studies on actual repositories to assess the practical utility of our method. 
We report results such as the number of accepted PRs and published CVEs, demonstrating the real-world impact of our approach.

\section{Results}
\label{sec:results}

\subsection{RQ1: How well does \method perform on the benchmark?}

\begin{table}[ht]
\centering
\caption{Performance comparison on the synthetic benchmark (\%). The bottom row shows percentage improvement (+) or decrease (-) compared to the next best baseline for each metric.}
\small
\setlength{\tabcolsep}{6pt}
\begin{tabular}{lcccccccc}
\toprule
\textbf{Model} & \textbf{P@1} & \textbf{P@3} &  \textbf{P@5} & \textbf{MAP} \\
\midrule
\rowcolor[HTML]{EFEFEF} 
Hash-based & \textbf{97.92} & 38.89 & 23.33 & 23.16 \\
ReDeBug & 95.74 & 39.72 & 23.83 & 23.26 \\
\rowcolor[HTML]{EFEFEF} 
MOVERY & 92.06 & 51.85 & 32.06 & 30.94 \\
FIRE & 94.23 & 41.03 & 24.62 & 24.10 \\
\rowcolor[HTML]{EFEFEF} 
SrcVul & 22.68 & 18.21 & 15.05 & 15.74 \\
\textbf{\method} & 94.85 &  \textbf{83.16} & \textbf{68.25} & \textbf{67.69} \\
\midrule
\rowcolor{gray!20}
\% Ours vs. Best & -3.14\% & +60.39\% & +112.88\% & +118.78\% \\
\bottomrule
\end{tabular}

\label{tab:rq1-results}
\end{table}

\subsubsection{Quantitative Analysis}
Table~\ref{tab:rq1-results} shows that \method outperforms all baseline methods across key evaluation metrics, showing strong retrieval accuracy and ranking quality.

Looking at the first prediction, \method achieves a P@1 of 94.85\%,
losing to the performance of the Hash-based method at 97.92\%. However, at larger retrieval depths, \method performs much better. 
Its P@3 reaches 83.16\%, clearly outperforming MOVERY, the best baseline at this depth, which scores 51.85\%. 
At the retrieval depth of $k = 5$,
\method achieves a P@5 of 68.25\%, more than doubling MOVERY's 32.06\%. 
Additionally, \method achieves a MAP of 67.69\%, representing a 119\% improvement over MOVERY’s MAP of 30.94\%, showing that \method ranks true positives much higher in the list.

Among the baselines, MOVERY performs best overall at deeper ranks with the highest P@3, P@5, and MAP. 
Hash-based shows the strongest early precision at P@1 of 97.92\%, while ReDeBug delivers balanced results across ranks.

\begin{figure}[h]
\centering
\includegraphics[width=0.8\textwidth]{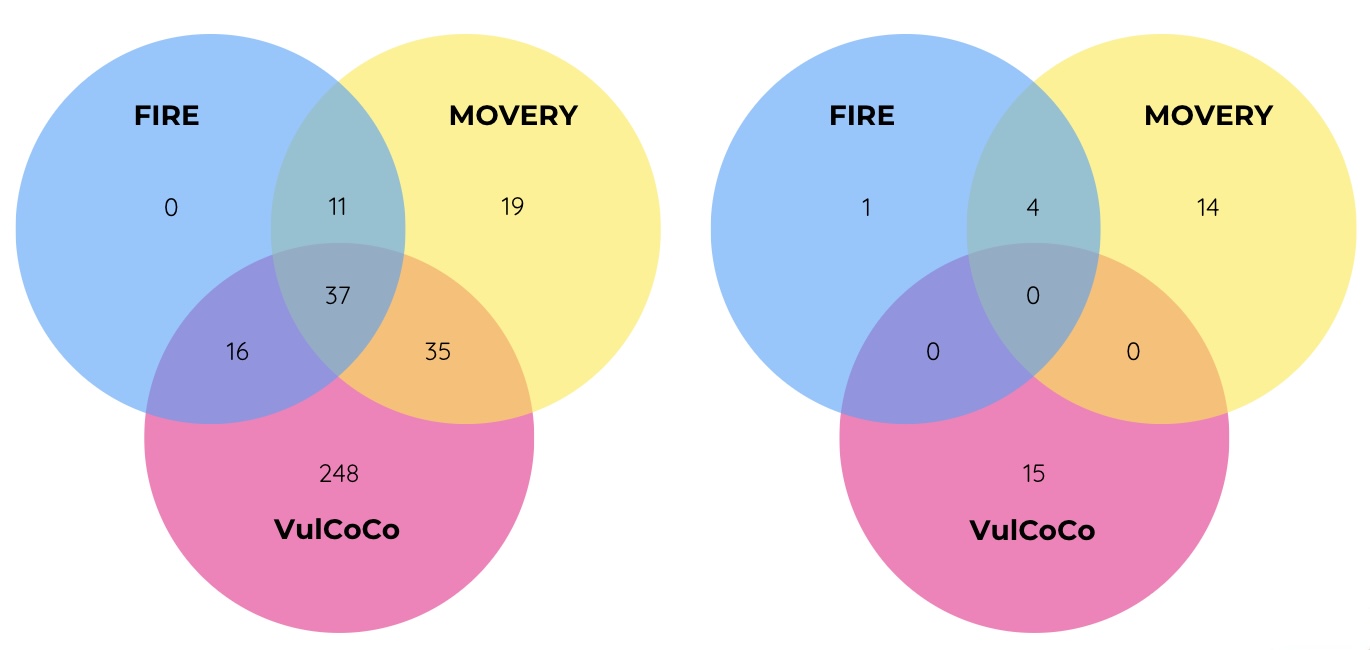}
\caption {Overlap of correct predictions (true positives, left) and wrong predictions (false positives, right) on the \benchmark benchmark.}
\label{fig:venns}
\end{figure}

To further examine prediction overlap, Figure~\ref{fig:venns} shows Venn diagrams. 
The left side illustrates true positives, which are the absolute number of true vulnerable clones correctly identified by different methods.
The right diagram shows the false positives, where non-vulnerable functions were incorrectly marked as vulnerable. 
Here, we focus on the three strongest techniques: FIRE, MOVERY, and \method. Notably, \method identifies the largest number of unique results (248 out of 336). 
MOVERY finds 19 unique cases, and FIRE cannot find any. 
Only 37 results are shared by all three methods, meaning most of \method's findings are not detected by others. This suggests that \method can find more subtle or difficult clones.

In contrast, the right side of the figure shows the false positives. 
\method produces 15 unique false positives, while MOVERY contributes 14, and FIRE generates 1. 
The limited overlap among different methods' failures suggests they rely on different heuristics or signals to make predictions.
Overall, \method's broader coverage does not come with a large increase in failures, showing that it maintains a good balance between precision and recall.

In conclusion, these results confirm that \method is highly effective at retrieving vulnerable clones, especially at deeper ranks.
The qualitative differences in both successes and failures highlight that \method offers a new and useful perspective.
It does not just copy what existing tools do, but instead adds a different way to find hidden and less obvious vulnerabilities.

\subsubsection{Qualitative Analysis}
Other than the quantitative analysis, we also conduct quality analysis to better understand what the advantages brought by \method and why it fails.
 
\vspace{4px}
\noindent\textbf{Failure Analysis for \method}
\method incorrectly classify 15 cases as VCCs when they are actually clones of fixed functions. Upon careful investigation of each case, we identify a common characteristic: the vulnerable and fixed versions of the original functions are nearly identical, differing by only 1-2 lines of code. The generated clones further blur these differences, causing \method to produce false positive predictions.

Among the cases that other tools successfully identified but \method failed to detect, 4 cases encountered JSON parsing errors due to invalid escape sequences. These errors occurred when unexpected characters in the response prevented correct parsing. For example, Listing~\ref{lst:json} shows Claude's response for one such case, which contains multiple backslash characters (\texttt{\textbackslash\textbackslash})  that complicate JSON processing.

\begin{lstlisting}[caption={Claude's response containing problematic escape characters}, label={lst:json},float=t]
{
  "is_vulnerable": true,
  "confidence_level": 5,
  "justification": "The cloned function contains the same SQL injection vulnerability as the original function. By comparing the original vulnerable function with the fixed version, the vulnerability was the lack of input validation for single quotes (') and backslashes (\\) in the timestring parameter, which could allow SQL injection attacks. The fix added validation: `if( strchr(timestring,'\'') || strchr(timestring, '\\') )` to reject malicious input. The cloned function, despite having obfuscated variable names (oler instead of timestring, uran instead of timefield, etc.), follows the exact same logic flow as the original vulnerable function but completely lacks the input validation check that was added in the fixed version. The cloned function directly processes the input parameter (oler, which corresponds to timestring) and passes it to createPostgresTimeCompareSimple and createPostgresTimeCompareRange functions without any sanitization, making it vulnerable to the same SQL injection attack vector."
}
\end{lstlisting}

For other cases, the original vulnerable functions and the corresponding generated clones are really long, with 63.70 lines on average while the original commit changes are really small, only 1-3 lines of code. This makes it difficult for \method to identify which specific lines contain the vulnerability when analyzing such large functions. The vulnerability fix is buried within many lines of unrelated code, making it hard for the model to focus on the few lines that actually matter for the security issue. 
Other tools like MOVERY~\cite{woo2022movery} and ReDeBug~\cite{jang2012redebug}, which focus specifically on the lines that were modified in vulnerability-fixing commits, are able to detect such cases.

\vspace{4px}
\noindent\textbf{Success analysis for \method where other baselines fail.}
Among the 248 cases where \method succeeds while other baselines fail, the distribution is 28 Type 1, 42 Type 2, 126 Type 3, and 52 Type 4 clones. The predominance of Type 3 and Type 4 successes highlights \method's strength in handling heavily transformed code. The embedding-based retrieval component enables \method to identify semantic similarities even when surface-level syntax has been extensively modified, while the LLM validation step can reason about vulnerability logic that persists despite structural changes. This two-stage approach proves particularly effective for Type 3-4 clones where traditional syntactic similarity measures used by baseline tools fail to match the clones to the vulnerable source functions.


\find{
Answer to RQ1: \method demonstrates superior performance on the synthetic benchmark, outperforming all baselines across the key evaluation metrics. 
It achieves the P@1 of 94.85\%, competitive with the best-performing baseline, while significantly outperforming the next best baseline by 60.39\% in P@3 and 112.88\% in P@5. With an MAP of 67.69\%, \method effectively ranks true vulnerable clones higher, making it well-suited for real-world scenarios where discovering multiple vulnerabilities across repositories is crucial.
}

\subsection{RQ2: How do different components affect \method?}

Although the final configuration of \method is selected based on validation results, here we report the performance of alternative settings on the test set for comparison. This ensures a fair analysis, as the best configuration was not chosen based on its performance on the testing data.

\subsubsection{Effect of LLM validation}
We compare the full \method pipeline using Claude Sonnet 4~\cite{claude} with two variants: one that removes LLM validation entirely, and another that uses GPT-4.0~\footnote{\url{https://openai.com/index/gpt-4-research/}} 
as the validator. As shown in Table~\ref{tab:ablation-llm}, removing the LLM validation stage results in clear performance drops, P@5 falls from 68.25\% to 63.09\%, and MAP drops from 67.69\% to 62.80\%. Using GPT-4.0 partially recovers performance, with P@5 reaching 63.62\% and MAP at 62.80\%. Claude Sonnet 4 consistently delivers stronger results across all retrieval depths, confirming the importance of high-quality LLM validation in improving precision@k and MAP.

\begin{table}[h]
\centering
\small

\label{tab:ablation-llm}
\setlength{\tabcolsep}{6pt}
\caption{Ablation study: Effect of LLM validation on benchmark performance. All metrics shown as percentages. The bottom row shows percentage improvement (+) or decrease (-) when replacing Claude Sonnet 4 with GPT-4.0.}
\begin{tabular}{lcccccccc}
\toprule
\textbf{Model} & \textbf{P@1} & \textbf{P@3} &  \textbf{P@5} & \textbf{MAP} \\
\midrule
\rowcolor[HTML]{EFEFEF} 
\method w/o LLM & 85.57 & 73.20 & 63.09 & 62.80 \\
Validation w/ GPT4.0    & \textbf{96.81} & 81.21 & 63.62 & 63.35 \\
\rowcolor[HTML]{EFEFEF} 
Validation w/ Claude Sonnet 4        & 94.85 & \textbf{83.16} & \textbf{68.25} & \textbf{67.69} \\
\midrule
\rowcolor{gray!20}
\% Difference & +2.06\% & -2.34\% & -6.78\% & -6.41\% \\
\bottomrule
\end{tabular}

\label{tab:ablation-llm}
\end{table}

\subsubsection{Effect of retrieval embedding model}
We assess the impact of the retrieval embedding model by replacing the default encoder in \method with alternative pre-trained models. 

We compare the following encoders:
\begin{itemize}[leftmargin=*]
    \item \texttt{jina-embeddings-v4}~\cite{günther2025jinaembeddingsv4universalembeddingsmultimodal}: Our default model, a universal embedding model designed for multimodal and multilingual retrieval. Built on \texttt{Qwen2.5-VL-3B-Instruct}~\cite{qwen2.5-VL}, it supports single and multi-vector retrieval across 30+ languages and various domains, including technical and visually complex documents. It includes task-specific adapters that can be selected at inference time, such as for retrieval and code understanding.
    \item \texttt{jina-embeddings-v3}~\cite{sturua2024jinaembeddingsv3multilingualembeddingstask}: The predecessor of \texttt{jina-embeddings-v4}, \texttt{jina-embeddings-v3} is a multilingual text embedding model based on XLM-RoBERTa with task-specific adapters for retrieval and similarity tasks. 
    \item \texttt{jina-embeddings-v2-base-code}~\cite{jinaai}: An earlier code-focused model in the Jina embeddings series, \texttt{jina-embeddings-v2-base-code} is trained on GitHub code and QA pairs across 30 programming languages.
    \item \texttt{Qwen3-4B}~\cite{qwen3technicalreport}: A multilingual language model trained for both general-purpose and coding tasks, with strong capabilities in reasoning, instruction-following, and agent-based interactions.
    \item \texttt{all-MiniLM-L6-v2}~\cite{MiniLM}: A small sentence embedding model trained on general textual similarity data.
\end{itemize}

While some models such as \texttt{all-MiniLM-L6-v2} and \texttt{jina-embeddings-v3} achieve higher P@1, they perform worse in overall MAP. Our default encoder, \texttt{jina-embeddings-v4}, achieves the best MAP at 67.69\% and P@5 at 68.25\%, demonstrating superior performance at retrieving relevant clones beyond the top result. 
This advantage in deeper ranking metrics is crucial for vulnerability detection, as it increases the pool of candidate clones available for LLM validation, ultimately reducing missed vulnerabilities.

\begin{table}[ht]
\centering
\caption{Ablation study: Effect of retrieval embedding model. All metrics shown as percentages. Bottom row shows percentage change when replacing \texttt{jina-embeddings-v4} with the next best performing model.}
\small
\label{tab:ablation-embed}
\setlength{\tabcolsep}{6pt}
\begin{tabular}{lcccccccc}
\toprule
\textbf{Model} & \textbf{P@1} & \textbf{P@3} &  \textbf{P@5} & \textbf{MAP} \\
\midrule
\rowcolor[HTML]{EFEFEF} 
\texttt{jina-embeddings-v4} & 94.85 & \textbf{83.16} & \textbf{68.25} & \textbf{67.69} \\
\texttt{jina-embeddings-v3} & 99.47 & 81.58 & 64.63 & 64.23 \\
\rowcolor[HTML]{EFEFEF} 
\texttt{jina-embeddings-v2-base-code} & 94.85 & 68.38 & 53.61 & 53.00 \\
\texttt{all-MiniLM-L6-v2} & \textbf{100.00} & 80.10 & 61.49 & 61.75 \\
\rowcolor[HTML]{EFEFEF} 
\texttt{Qwen3-4B} & 93.48 & 76.09 & 58.26 & 57.41 \\
\midrule
\rowcolor{gray!20}
\% Difference & +2.16\% & -5.87\% & -10.50\% & -10.63\% \\
\bottomrule
\end{tabular}

\end{table}

\subsubsection{Effect of similarity threshold $t$}
We examine how the similarity threshold $t$, used to filter candidates based on embedding similarity, affects retrieval performance. A higher threshold $t$ yields fewer but more confidently similar candidates, improving precision at the cost of recall. In contrast, lowering $t$ expands the candidate set and may improve recall but risks introducing false positives. 

As shown in Table~\ref{tab:ablation-threshold}, retrieval performance improves consistently as the threshold is lowered from 0.9 to 0.7. In particular, MAP rises from 50.54\% to 67.69\%, and P@3 improves from 72.28\% to 83.16\%. This reflects a significant increase in the number of relevant clones ranked higher when the candidate pool is broadened. The gains from 0.8 to 0.7 are smaller but still present, while scores remain unchanged entirely between 0.7 and 0.6. This suggests that $t = 0.7$ provides an effective trade-off: it admits a sufficiently broad candidate set to recall true positives while keeping false positives low enough for downstream filtering.

\begin{table}[ht]
\centering
\caption{Effect of similarity threshold $t$. All metrics shown as percentages. Bottom row shows percentage change when changing this threshold from 0.7 to 0.9.}
\small
\setlength{\tabcolsep}{6pt}
\begin{tabular}{lcccccccc}
\toprule
\textbf{Threshold $t$} & \textbf{P@1} & \textbf{P@3} & \textbf{P@5} & \textbf{MAP} \\
\midrule
\rowcolor[HTML]{EFEFEF} 
0.9 & 94.74 & 72.28 & 51.37 & 50.54 \\
0.8 & 94.85 & 82.82 & 67.63 & 67.07 \\
\rowcolor[HTML]{EFEFEF} 
0.7 & 94.85 & 83.16 & 68.25 & 67.69 \\
0.6 & 94.85 & 83.16 & 68.25 & 67.69 \\
\midrule
\rowcolor{gray!20}
\% Difference (0.7 vs. 0.9) & -0.12\% & -13.08\% & -24.73\% & -25.34\% \\

\bottomrule
\end{tabular}

\label{tab:ablation-threshold}
\end{table}

\find{
Answer to RQ2: The ablation study reveals that each component of \method contributes significantly to the overall performance. 
The LLM validation step is crucial, with Claude Sonnet 4 outperforming GPT-4.0 and providing improvements over no validation with P@5 of 68.25\% from 63.09\%. 
For the embedding model, \texttt{jina-embeddings-v4} achieves the best overall performance with highest MAP of 67.69\% and P@5 of 68.25\%, demonstrating strong performance in identifying VCCs across multiple depths. A similarity threshold of 0.7 provides the optimal effectiveness, with MAP improving from 50.54\% to 67.69\% when this threshold is lowered from 0.9. 
}

\subsection{RQ3: How effective is \method in the real world scenario?}
\subsubsection{Running \method on real-world projects.}

We run \method and the baselines using known vulnerable fixing commits
in PrimeVul~\cite{ding2024vulnerability} and CleanVul~\cite{li2024cleanvul} as queries to identify VCCs in real-world projects. 
For tools that operate at the function level (e.g., Hash-based, MOVERY, FIRE, SrcVul, and \method), we use function-level data in the datasets to align with their expected input format. 

While we use a similarity threshold of 0.7 in the benchmark setting to evaluate the upper bound of clone retrieval, we increase the threshold to 0.85 in the real-world evaluation. A lower threshold yields more candidate clones, but also results in a much larger number of candidates for the LLM to validate and substantially increasing financial cost. To strike a balance between detection coverage and financial cost, we choose a more strict threshold of 0.85 in the real-world setting, which reduced the validation load while still maintaining competitive detection performance. Table~\ref{tab:realworld} presents the effectiveness of each approach in terms of true/false positives, false negatives, precision, recall, and F1 score.

\method successfully detects all 127 vulnerable clones with a precision of 56.44 and a perfect recall, achieving the best F1 score of 72.16\%. It detects every clone found by the other tools and also finds some that the others miss. This shows that \method has a strong coverage while still keeping a decent precision. In contrast, FIRE has a perfect precision but only finds 17 vulnerable clones, which gives it a recall of 13.39\% and an F1 score of 23.61\%. MOVERY and ReDeBug report even fewer true positives with low recall scores of 8.61\% and 14.17\%, leading to F1 scores of just 14.67\% and 21.30\%, respectively. On the other hand, SrcVul returns an extremely large number of positives, in the millions, which mostly should be false positives. To confirm this, we randomly sample 100 positive clones from SrcVul and find that none of them were vulnerable clones. The hash-based method fails to produce any predictions in the real-world setting.

In summary, \method significantly outperforms all baselines in real-world vulnerability clone detection, especially in recall. Compared to FIRE, which is the best-performing baseline, it improves recall by 647\% and F1 by 206\%, while maintaining high precision. These results demonstrate the robustness of \method in realistic scenarios where clone variants often differ in structure or formatting.

\begin{table}[ht]
\centering
\caption{Real-world evaluation results: TP = true positives, FP = false positives, FN = false negatives.}
\small
\setlength{\tabcolsep}{6pt}
\begin{tabular}{lcccccc}
\toprule
\textbf{Model} & \textbf{TP} & \textbf{FP} & \textbf{FN} & \textbf{Precision} & \textbf{Recall} & \textbf{F1} \\
\midrule
\rowcolor[HTML]{EFEFEF} 
Hash-based & 0 & 0 & 127 & 0.00 & 0.00 & 0.00 \\
ReDeBug & 18 & 24 & 109 & 42.86 & 14.17 & 21.30 \\
\rowcolor[HTML]{EFEFEF} 
MOVERY & 11 & 12 & 116 & 47.83 & 8.61 & 14.67 \\
FIRE & 17 & 0 & 110 & 100.00 & 13.39 & 23.61 \\
\rowcolor[HTML]{EFEFEF} 
SrcVul & 0 & 100 & 127 & 0.00 & 0.00 & 0.00 \\
\midrule
\textbf{\method} & 127 & 98 & 0 & 56.44 & 100.00 & 72.16 \\
\bottomrule
\end{tabular}

\label{tab:realworld}
\end{table}

\paragraph{False Positive Analysis for \method.} 
In one false positive case, our method retrieved the function \texttt{h2v2\_fancy\_upsample} from the \texttt{aseprite/aseprite}\footnote{\url{https://github.com/aseprite/aseprite}} repository. 
As shown in Listing~\ref{lst:h2v2}, \method flagged this function as vulnerable due to potentially unsafe array accesses.

\begin{figure}[H]
\centering
\begin{lstlisting}[caption={Snippet from \texttt{h2v2\_fancy\_upsample}}, label={lst:h2v2}]
METHODDEF(void)
h2v2_fancy_upsample (j_decompress_ptr cinfo, jpeg_component_info * compptr,
                     JSAMPARRAY input_data, JSAMPARRAY * output_data_ptr)
{
...
if (v == 0)
  inptr1 = input_data[inrow-1];
else
  inptr1 = input_data[inrow+1];

outptr = output_data[outrow++];
...
\end{lstlisting}
\end{figure}

Although the LLM justification presents a plausible vulnerability, the prediction is considered a false positive because it does not match the actual vulnerability fix for the source function \texttt{merged\_2v\_upsample}. As shown in Listing~\ref{lst:fix}, the original fix changed the type from \texttt{my\_upsample\_ptr} to \texttt{my\_merged\_upsample\_ptr}.

\begin{figure}[H]
\centering
\begin{lstlisting}[caption={Patch in \texttt{merged\_2v\_upsample}}, label={lst:fix}]
  METHODDEF(void)
  merged_2v_upsample(j_decompress_ptr cinfo, ...)
  {
- my_upsample_ptr upsample = (my_upsample_ptr)cinfo->upsample;
+ my_merged_upsample_ptr upsample = (my_merged_upsample_ptr)cinfo->upsample;
  JSAMPROW work_ptrs[2];
  JDIMENSION num_rows;
  ...
\end{lstlisting}
\end{figure}

This modification corrects the use of an erroneous internal data structure, which may have caused segmentation faults during execution. The vulnerability is fundamentally different from array bounds issues.

This case illustrates two compounding issues in function-level vulnerable clone detection. First, operating at the function level, the simple type change from \texttt{my\_upsample\_ptr} to \texttt{my\_merged\_upsample\_ptr} may not appear as an obvious vulnerability fix. It looks like a simple refactoring rather than a security patch. Without broader context about structure layout mismatches and their implications, the actual vulnerability is essentially invisible at the function scope.

Second, when the LLM cannot identify the true vulnerability that was fixed, it exhibits vulnerability confabulation~\cite{sui2024confabulation}—a tendency to justify the function as vulnerable by focusing on other potentially risky patterns in the code. Rather than acknowledging uncertainty, the LLM confidently identifies array indexing patterns as the vulnerability, creating a plausible but incorrect justification. This pattern is representative of other false positive cases we observed, highlighting the inherent difficulty of function-level vulnerability detection when dealing with subtle, context-dependent security issues.


\subsubsection{Submitting pull requests on detected clones.}

To evaluate the practical usage of our clone detection pipeline, we submitted security PRs to open-source projects hosting vulnerable clones identified by our system. Over the course of our study, we submitted a total of 400 PRs to 284 open-source repositories, each referencing a known vulnerability, typically linked to a CVE, and its corresponding fix from another project. These PRs targeted repositories that replicated vulnerable code, often without awareness of its security implications.

Each submission included a concise explanation of the issue, a reference to the original vulnerability fixing commit, and a proposed fix for the clone. In some cases, we adapted the fix to suit the specific context of the host project. This process demonstrated the practical value of our approach in discovering VCCs that might otherwise go unnoticed.

As a direct result of these efforts, 75 PRs were merged to the main branch, and 15 CVEs were published 
either newly assigned to the affected projects or publicly acknowledged by maintainers. We provide the full list of these CVEs in our replication package\footnote{\url{https://github.com/tabudz/VulCoCo}}, which includes details such as project names, CVE identifiers, and corresponding patches. This confirms that VCCs in the wild often remain unpatched until discovered by targeted analysis and community engagement. Our submissions not only helped secure a number of open-source projects but also contributed new entries to the vulnerability knowledge base.

On the other hand, 89 PRs were closed for various reasons. The majority of these cases involved correctly detected VCCs in functions that target projects do not actually use, making the fixes unnecessary. Several authors requested proof that their systems were actually vulnerable, which is not trivial to provide. Some maintainers acknowledged the vulnerabilities but chose to implement their own fixes rather than accept our patches. There are also PRs which were closed without any explanation.

These results highlight both the promise and practical challenges of automated vulnerability discovery. The 75 merged PRs and 15 published CVEs confirm the real-world value of clone-based detection, while the 89 closed PRs show the importance of understanding project contexts and maintainer expectations. Overall, our approach demonstrates a concrete pathway from research to security impact, though success depends on careful community interaction.

\find{
Answer to RQ3: Our real-world evaluation shows that \method significantly outperforms existing methods in practical VCC detection scenarios. 
\method outperforms the best baseline FIRE by 23.61\% in terms of F1 score.
We also submit 400 PRs to 284 open-source projects with the help of \method, leading to 15 newly published CVEs.
}
\section{Discussion}
\label{sec:discussion}

\subsection{Threats to Validity}

\noindent\textbf{Threats to internal validity.}
A potential threat to internal validity is ensuring the fair and correct implementation of baseline methods. 
To mitigate this, we utilized the official replication packages provided by the original authors for all baselines. 
This ensures each method was executed in its intended environment with the prescribed configurations, guaranteeing a fair comparison.


\vspace{4px}
\noindent\textbf{Threats to external validity.} 
The primary threat to external validity concerns whether our findings can be generalized beyond the languages used in our evaluation. 
Our benchmark is composed of C/C++ functions, which could limit the applicability of our results. 
To address this, we conducted an additional evaluation of \method on real-world Java repositories. 
The experimental results demonstrated that our method is effective in this different context, which provides evidence that its underlying principles are not language-specific. 
We acknowledge that further experiments on other languages would be necessary to fully establish broad generalizability, but our cross-language validation significantly mitigates this threat.


\vspace{4px}
\noindent\textbf{Threats to construct validity.}
The first threat to construct validity is the subjectivity inherent in our manual validation of the synthetic dataset and the retrieved VCCs. 
To mitigate this, the labeling was performed independently by two authors, each with over 5 years of programming experience. 
Crucially, this process was conducted blindly; the labelers were unaware of which method produced the VCCs, which minimizes potential bias. 
We therefore believe this threat is well-controlled.

The second threat relates to the appropriateness of the evaluation metrics. 
When evaluating the methods on the benchmark, we employ standard retrieval metrics like Precision@k and MAP. 
When evaluating the methods in real-world projects, we follow prior studies to use precision, recall, and F1 score.
Since these metrics are widely used by prior VCC detection methods~\cite{woo2022movery,kim2017vuddy,feng2024fire}, we consider the threats to be minor.


\subsection{Vulnerability Disclosure}
When we found VCCs in open-source projects, we followed the responsible disclosure protocol recommended by each targeted project. We first checked the project's \texttt{CONTRIBUTING.md} or \texttt{SECURITY.md} to see their preferred way to report security issues. Based on this, we either submitted a public PR on GitHub or contacted the maintainers through private email. In many cases, project maintainers appreciated our reports and accepted our PRs with fixes. These contributions helped improve the security of their projects and even led to the publication of new CVEs.

However, not all maintainers responded the same way. Some were not very concerned about the issues we reported, or they thought that the code did not resemble the original vulnerability, and they required a full proof-of-concept to exploit their project. In such cases, our fixes were rejected or ignored. These experiences reflect the challenges of responsible vulnerability disclosure, especially when the bugs are subtle or lack obvious exploitability.

\section{Related Work}
\label{sec:related}
In this section, we primarily discuss the three research lines that are most relevant to our work: general code clone detection, vulnerability detection in code, and vulnerable clone detection. Our method builds upon and differentiates from these efforts by combining semantic retrieval with LLM-guided validation and introducing a controlled benchmark for evaluation.

\subsection{General Code Clone Detection}

Code clone detection has been extensively studied, with techniques broadly categorized into text-based, token-based, AST-based, and semantic approaches~\cite{roy2009comparison}. 
Traditional tools such as NiCad~\cite{cordy2011nicad}, Deckard~\cite{jiang2007deckard}, and SourcererCC~\cite{sajnani2016sourcerercc} focus on identifying syntactic or structural similarity. While effective for general clone detection, these approaches struggle with minor edits or semantic changes, and they are not tailored to detect security vulnerabilities in clones.

Recent work has shifted toward deep learning-based methods, with pretrained models like CodeBERT~\cite{feng2020codebert}, GraphCodeBERT~\cite{guo2020graphcodebert}, and CodeT5~\cite{wang2021codet5} capturing deeper semantic features of code. These embeddings have been evaluated on clone detection benchmarks~\cite{lu2021codexglue}.

However, these general-purpose models are not optimized for security contexts: they lack vulnerability-specific treatment or downstream tasks like clone-based vulnerability detection. In contrast, our method leverages semantic retrieval with such models but adds a critical LLM validation step, which substantially improves precision in identifying true vulnerable clones rather than merely similar code.

\subsection{Vulnerability Detection in Code}

A significant body of work has focused on detecting vulnerabilities directly in source code, leveraging both static analysis and machine learning techniques. Traditional approaches like SySeVR~\cite{li2021sysevr} and VulDeePecker~\cite{li2018vuldeepecker} identify vulnerability-prone code patterns using static slicing techniques coupled with deep neural models, which automatically learn vulnerability-relevant features from code gadgets. Devign~\cite{zhou2019devign}, for example, models control and data dependencies using graph neural networks to predict whether a function contains a vulnerability.

These systems operate on individual functions or code blocks and return a binary vulnerability label or probability score. However, they do not leverage any known vulnerability as reference and typically lack contextual justification for the prediction. This can limit their usefulness in practice, where developers may require evidence to trust or act on automated results.

In contrast, our method is designed for a more focused setting (i.e., clone-based vulnerability detection) where the goal is to find semantically similar code that replicates a known vulnerability. Starting from a reference vulnerable function, our method retrieves candidate clones and validates them using a commercial LLM, which provides both a classification and a natural language explanation of whether the vulnerability is still present. This enables not only higher precision but also supports downstream tasks such as automated pull request generation and CVE discovery with interpretable evidence.

\subsection{Vulnerable Clone Detection}
Vulnerable clone detection targets the discovery of code clones that retain the vulnerability of a known flawed implementation. Early tools like ReDeBug~\cite{jang2012redebug} and VUDDY~\cite{kim2017vuddy} rely on syntactic matching and context-based heuristics to detect clones of known vulnerable code. While scalable, these tools struggle when clones deviate from the original form through semantic refactoring.

To address these limitations, MVP~\cite{xiao2020mvp} introduces patch-enhanced vulnerability signatures that combine slicing and differential analysis of vulnerable and fixed versions. This allows MVP to distinguish between clones that remain vulnerable and those that have been patched. SrcVul~\cite{alomari2025slicing} further improves accuracy by extracting vulnerability-related slices around key variables and applying locality-sensitive hashing for matching.

Our method differs in two key ways. First, it integrates an LLM-guided validation step to assess whether a candidate clone still retains the core vulnerability—beyond syntactic or slice-based similarity. Second, we contribute a synthetic benchmark designed to simulate realistic perturbations, enabling controlled evaluation across clone types and retrieval depths. Unlike prior work that often evaluates only on real-world case studies, our benchmark provides a more systematic and reproducible testbed for clone-based vulnerability detection methods.

\section{Conclusion and Future Work}
\label{sec:conclusion}
In this paper, we present a novel framework \method for detecting VCCs by retrieving semantically similar functions to known vulnerabilities and validating them using an LLM. 
Unlike many prior tools that only return a binary label or a probability score indicating whether a piece of code is vulnerable, our method provides a more informative and transparent result. 

To support reproducible evaluation, we first build a synthetic benchmark named \benchmark that includes 1,000 synthesized vulnerable and non-vulnerable functions.
\method outperforms prior state-of-the-art methods.
When tested on the \benchmark benchmark, \method achieves up to 119\% higher MAP compared to the best-performing baseline, i.e., MOVERY~\cite{woo2022movery}.
When tested on real software projects, \method can also find more than 7 times as many VCCs compared to the best-performing method, i.e., FIRE~\cite{feng2024fire}, while keeping good precision.


In addition, to further evaluate our system's practical impact, we submitted 400 PRs to open-source repositories based on detected clones. 
Of which, 75 of them are merged, while 89 PRs are closed with various reason, and the rest received no response. 
Importantly, not all closed PRs represent false positives, several maintainers acknowledged the issues as genuine vulnerabilities but chose alternative solutions.
This effort resulted in the publication of 15 CVEs, highlighting the real-world applicability and effectiveness of our approach.

In future work, we plan to explore several directions. 
First, we aim to expand the scope of source vulnerabilities by automatically curating a more diverse and comprehensive vulnerability corpus across programming languages and vulnerability types.
Second, we also intend to explore lightweight alternatives, such as fine-tuned models or hybrid rule-based checks, to improve the reliability and cost-efficiency of the LLM validation stage.
Third, we are interested in extending our framework to support automated patch suggestion, closing the loop from vulnerability discovery to repair. 
Finally, we hope to collaborate with maintainers and security teams to better understand the social and technical challenges of integrating clone-based vulnerability detection into secure software development pipelines.

\newpage
\bibliographystyle{ACM-Reference-Format}
\bibliography{main}
\end{document}